\documentclass[fleqn,twoside]{article}
\usepackage{espcrc2,epsfig}
\title{Phenomenology of Neutrino Oscillations}
\author{G.L.\ Fogli\address[BARI]{Dipartimento di Fisica
        and Sezione INFN di Bari, Via Amendola 173, 70126 Bari, Italy},
    E.\ Lisi\addressmark[BARI]\thanks{Speaker. E-mail:\ \tt
        eligio.lisi@ba.infn.it},
    A.\ Marrone\addressmark[BARI],
    D.\ Montanino\address[LECCE]{Dipartimento di Fisica
        and Sezione INFN di Lecce, Via Arnesano, 73100 Lecce, Italy},
    and A.\ Palazzo\addressmark[BARI]}
\begin{document}
\begin{abstract}
We review the status of several phenomenological topics of current
interest in neutrino oscillations: (i) Solar neutrino oscillations
after the first Sudbury Neutrino Observatory measurements,
including both model-independent and model-dependent results; (ii)
Dominant $\nu_\mu\to\nu_\tau$ oscillations of atmospheric and K2K
neutrinos,  and possible subdominant oscillations induced by
either extra states or extra interactions; and (iii) Four-neutrino
scenarios embedding the controversial LSND evidence for
oscillations. \vspace{0pc}
\end{abstract}
\maketitle

\section{INTRODUCTION}

Evidence in favor of $\nu$ oscillations is  accumulating at a fast
rate. The strong evidence for  disappearance of atmospheric
$\nu_\mu$ in Super-Kamiokande (SK) \cite{Tots}, consistent with
the $\mu$ spectrum anomaly in MACRO \cite{Scap} and with Soudan-2
 $(\mu,e)$ data \cite{Pety}, is successfully explained by
$\nu_\mu\to\nu_\tau$ oscillations. The recent data from the first
long-baseline K2K experiment \cite{Hase} also support such an
interpretation \cite{K2Ko}. The global evidence for solar $\nu_e$
oscillations from radiochemical (Cl and Ga) \cite{Catt,Verm} and
Cherenkov (SK and SNO) detectors \cite{McDo} is strongly
corroborated (at a significance level $>3\sigma$ \cite{Bah4}) by
the comparison of the recent SNO data \cite{SNOp,Laws} with those
collected in SK \cite{SKso}.

We review ``established'' aspects of the above pieces of evidence,
as well as (partially) ``unsolved'' issues concerning sterile
neutrinos, new neutrino interactions, and the controversial
$\nu_\mu\to\nu_e$ signal in LSND \cite{LSND}.

\section{SOLAR NEUTRINOS AFTER SNO}

The main goal of the SNO heavy-water experiment is to measure the
solar neutrino flux both in charged current mode
$$
\nu_e + d \to p + p + e^- \ ({\rm CC})\ ,
$$
sensitive to $\nu_e$ only, and in neutral current mode
$$
\nu_x + d \to p + n + \nu_x \ ({\rm CC})\ ,
$$
sensitive to any active flavor $\nu_x=\nu_{e,\mu,\tau}$. A
suppression of the CC/NC ratio is expected in the presence of
$\nu_e\to\nu_x$ oscillations. The first CC data have recently been
released \cite{SNOp}, while the NC measurement is still in
progress \cite{McDo}.

The SK experiment is also sensitive to CC and NC reactions through
elastic scattering on $e^-$,
$$
\nu_x + e^- \to \nu_x + e^- \ ({\rm ES})\ ,
$$
but no discrimination is possible for the event type (CC or NC).
However, one can indirectly use the SNO CC measurement  to
``extract'' a possible NC component in SK. Indeed, this component
turns out to be nonzero at $>3\sigma$ level, indicating that
$\nu_e\to\nu_{\mu,\tau}$ transitions are occurring.

\subsection{Model-independent implications}

Assuming active $\nu$ oscillations, the main unknowns for the
calculations of the SNO CC rate are the $\nu_e$ survival
probability averaged over the detector response function, $\langle
P_{ee}\rangle_{\rm SNO}$, and the ratio $f_B$ of the absolute
$^8$B spectrum to the reference one estimated in the Standard
Solar Model (SSM) of \cite{BP00}. Analogously, the unknowns
associated to the SK ES rate are $\langle P_{ee}\rangle_{\rm SK}$
and $f_B$. The two available data (SNO CC and SK ES rates) are
insufficient to pin down  three unknowns [$\langle
P_{ee}\rangle_{\rm SNO},\, \langle P_{ee}\rangle_{\rm SK},\,f_B$],
unless one make some hypothesis on either the SSM (e.g., by
setting $f_B=1$), or the oscillation model [e.g., by setting a
functional form of $P_{ee}(E_\nu)$)].

However, it has been shown in \cite{Vill,ViPa} that, by
appropriately shifting the SK energy threshold, one can reach an
accurate ``equalization'' of the SK and SNO response as a function
of the neutrino energy. For the current SNO threshold in electron
kinetic energy (6.75 MeV), the SK threshold providing the best
equalization  is 8.60 MeV, as shown in Fig.~1 \cite{SNOo}.

This lucky circumstance implies that the SNO CC and SK ES
experiments, above such thresholds, are sensitive to the {\em
same\/} energy-averaged survival probability $P_{ee}$,
$$
\langle P_{ee}\rangle_{\rm SNO}=\langle P_{ee}\rangle_{\rm
SK}\equiv\langle P_{ee}\rangle\ ,
$$
allowing a reduction to only two unknowns ($\langle
P_{ee}\rangle,\,f_B)$. The corresponding constraints placed by the
current SK and SNO  data are shown in Fig.~2 \cite{SNOo}. The data
prefer $f_B\sim1$, in agreement with the SSM expectation of
\cite{BP00}, and clearly indicate that $\langle P_{ee}\rangle < 1$
with a
significance greater than $3\sigma$.%
\footnote{ Within $\pm3\sigma$, it is $f_B=1.03^{+0.50}_{-0.58}$
and $\langle P_{ee}\rangle = 0.34^{+0.61}_{-0.18}$.}
Such conclusions are strictly model-independent: no a priori
assumption has been made on the SSM (i.e., on $f_B$) or on the
energy dependence of $P_{ee}(E_\nu)$.

\begin{figure}[t]
\vspace{-.7truecm} \hspace{-.5truecm}
\epsfig{bbllx=0pt,bblly=0pt,bburx=500pt,bbury=750pt,width=7truecm,%
figure=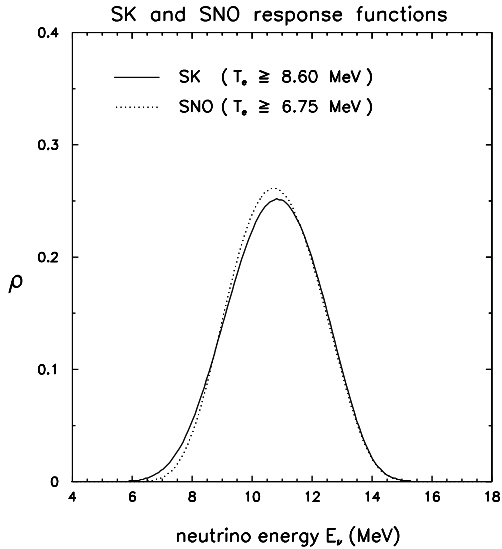} \vspace{-3truecm} \caption{\footnotesize SNO
response function for neutrino-induced CC events with $T_e>6.75$
MeV (dotted curve), and equalized SK response function for ES
events with $T_e>8.60$ MeV (solid curve) \protect\cite{SNOo}.
\vskip-6mm}
\end{figure}

\begin{figure}[t]
\vspace{-.7truecm} \hspace{-.5truecm}
\epsfig{bbllx=0pt,bblly=0pt,bburx=500pt,bbury=750pt,width=7truecm,%
figure=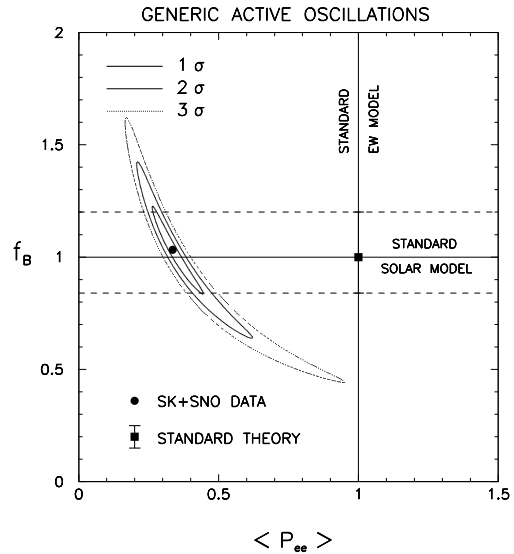} \vspace{-3truecm} \caption{\footnotesize SK+SNO
combined constraints on $f_B$ and $\langle P_{ee}\rangle$.
Preferred values are $f_B\sim 1$ (in agreement with the standard
solar model) and $P_{ee}\sim 1/3$ (inconsistent with the
hypothesis of no oscillations) \protect\cite{SNOo}. \vskip-6mm}
\end{figure}

\subsection{Active $\nu$ oscillations}

Since the model-independent analysis suggests the validity of both
the SSM and the oscillation hypothesis, it makes sense {\em to
assume\/} the SSM $\nu$ fluxes as input, and to  calculate the
oscillation probability in the simplest case of $2\nu$
oscillations (described by the mass-mixing parameters $\delta
m^2=m^2_2-m^2_1$ and $\omega=\theta_{12}$). The comparison with
the available solar neutrino data gives, through a $\chi^2$
statistics, allowed regions in the plane $(\delta
m^2,\tan^2\omega)$. In the analysis, it is useful to include also
CHOOZ reactor data \cite{CHOO}, which are relevant to bound high
values of $\delta m^2$.

Figure~1 shows the results of a global pre-SNO data analysis
\cite{SNOo}. There are several (almost) disconnected allowed
regions  in the mass-mixing plane, usually called---in the solar
$\nu$ jargon---as: Small Mixing Angle (SMA) at $\delta
m^2\sim10^{-5}$ eV$^2$ and $\tan^2\omega\sim 10^{-3}$; Large
Mixing Angle (LMA) at $\delta m^2\sim 10^{-4}$ eV$^2$ and
$\tan^2\omega\sim O(1)$; Low $\delta m^2$ (LOW) at $\delta m^2\sim
10^{-7}$ eV$^2$ and $\tan^2\omega\sim 1$; and multiple  Vacuum
Oscillation  (VO) solutions at large mixing and $\delta m^2\sim
O(10^{-9})$ eV$^2$, possibly connected to the the LOW solutions
through intermediate Quasi Vacuum Oscillation (QVO) solutions at
$\delta m^2\sim O(10^{-8})$ eV$^2$. The vertical ``gap'' at large
mixing for $\delta m^2\sim 10^{-6}$--$10^{-5}$ eV$^2$ is mainly
due to nonobservation of Earth matter effects  at SK \cite{SKso}.
The horizontal gap between the SMA and the LMA regions is kept
stable by SK spectral data.

\begin{figure}[t]
\vspace{-.truecm}
\hspace{-1.2truecm}
\epsfig{bbllx=0pt,bblly=0pt,bburx=500pt,bbury=750pt,width=8truecm,%
height=9truecm,figure=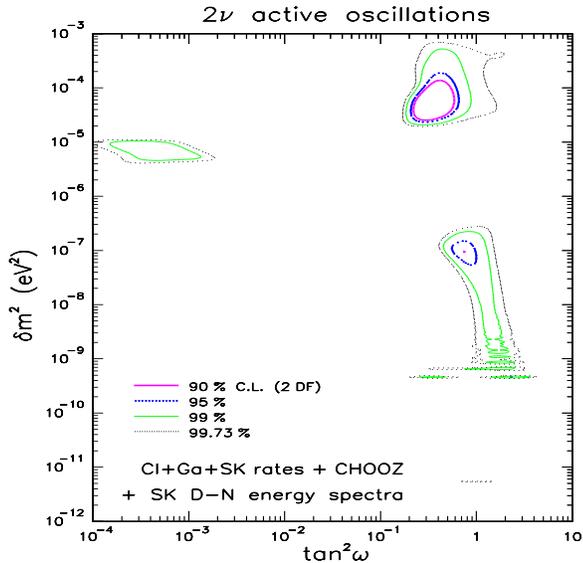} \vspace{-2.1truecm}
\caption{\footnotesize Global analysis of pre-SNO solar neutrino
data: Favored regions in the plane of the $2\nu$ mass-mixing
oscillation parameters \protect\cite{SNOo}. \vskip-7mm}
\end{figure}

Notice that, although pre-SNO data cannot exclude small mixing
cases (SMA), they globally show a preference for large mixing
$[\tan^2\omega\sim O(1)])$, mainly driven by SK spectral data
\cite{SKso,SNOo}. The preference for large mixing is strengthened
in the post-SNO analysis, as shown in Fig.~4. Indeed, the SMA
solution disappears at $\sim 3\sigma$ (while the LMA one provides
the best fit): a very important consequence of the first SNO CC
data. The reason is the following: The SMA solution in Fig.~3
typically predicts values of $\langle P_{ee}\rangle$ larger than
those favored in Fig.~2. In order to adapt to relatively low
values of $P_{ee}$, the SMA solutions tends to move rightwards,
where the nonadiabatic $\nu_e$ suppression is stronger and energy
spectrum distortions are larger; in doing so, however, the SMA
solution gets in conflict with the  nonobservation of spectral
distortions in SK, and becomes strongly disfavored in the global
fit.

Similar results have been obtained in \cite{Simi}. The extension
to $3\nu$ oscillations, within phenomenological bounds, does not
significantly alter the emerging post-SNO global picture
\cite{Fogl}.

\begin{figure}[t]
\vspace{-.truecm}
\hspace{-1.2truecm}
\epsfig{bbllx=0pt,bblly=0pt,bburx=500pt,bbury=750pt,width=8truecm,%
height=9truecm,figure=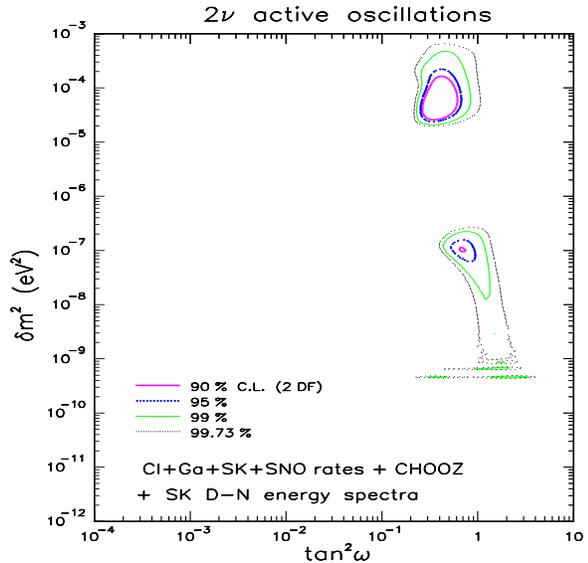} \vspace{-2.1truecm}
\caption{\footnotesize Global analysis of post-SNO solar neutrino
data \protect\cite{SNOo}. Notice that only solutions at large
mixing survive.\vskip-6mm}
\end{figure}

The surviving large-mixing solutions in Fig.~4 will soon be tested
in upcoming experiments. In particular, KamLand \cite{Scho} will
probe the LMA solution through reactor $\nu$ disappearance over
long baselines, while the  Borexino solar $\nu$ experiment
\cite{Shut,Anto} will probe the LOW and (Q)VO solutions through
day-night and seasonal variation effects, respectively. A
confirmation of the current best-fit solution (LMA) would be
extremely important for the physics potential of planned $\nu$
factory \cite{Dyda} or superbeam \cite{Koba} projects.

\subsection{Fate of the sterile neutrino}

The SNO-SK data comparison provides model-independent evidence for
$\nu_e$ oscillations into active species \cite{McDo,SNOp}. Can one
exclude additional transitions of solar $\nu_e$ to a sterile state
$\nu_s$?

The analysis in \cite{Barg} seems to show that a large $\nu_s$
component can be tolerated, provided that $f_B$ is significantly
increased: then $\nu_e\to\nu_s$ oscillations would make the extra
B $\nu$ flux unobservable in SNO and SK. However, the analysis in
\cite{Gonz} shows that, even with unconstrained $^8$B neutrino
flux, meaningful upper limits can be put on the $\nu_s$ admixture,
since for large-amplitude  $\nu_e\to\nu_s$ oscillations the global
fit becomes always worse (see Fig.~5). The difference in the
results can be traced \cite{Gonz} to the role of SK day-night
spectral data, which are used in \cite{Gonz} but not in
\cite{Barg}.

In any case, it can be safely concluded that there is no evidence
in favor of additional $\nu_s$ mixing, although subdominant
$\nu_e\to\nu_s$ transitions cannot be excluded within the still
large uncertainties. The limiting case of pure $\nu_e\to\nu_s$ is
strongly disfavored after SNO: accepting this case would be
equivalent to assume the the SNO-SK rate difference is a mere
$\sim3\sigma$ stat.\ fluctuation.

\begin{figure}
\psfig{width=7.5truecm,file=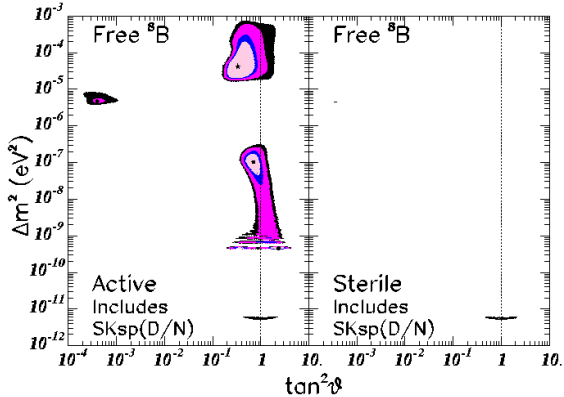} \vspace*{-1cm}
\caption{\footnotesize Comparison of solutions for $\nu_e$
oscillations into active states (left) or sterile state (right),
with free $f_B$. The active case provides a much better fit to the
data \protect\cite{Gonz}. \vskip-6mm}
\end{figure}

\section{ATMOSPHERIC NEUTRINOS}

The evidence for dominant $\nu_\mu\to\nu_\tau$ oscillations in
atmospheric neutrinos is robust. Figure~6 shows that the
neutrino-induced lepton events in SK, spanning 4 orders of
magnitude in energy $E$ and three orders of magnitude in $\nu$
pathlength $L$, are perfectly described by the simple hypothesis
of $2\nu$ oscillations. The best-fit mass-mixing parameters,
within a factor of about two, are stable around
$m^2=m^2_3-m^2_{1,2}\equiv m^2\simeq 3\times 10^{-3}$ and
$\tan^2\psi\equiv\tan^2\theta_{23}\simeq 1$ \cite{Tots,Subd}.

However, it is important to keep in mind that two important pieces
of information are still ``hidden'' in the data: A clear
observation of $\nu_\tau$ appearance (for which SK can only
provide interesting---but not really decisive---hints \cite{Tots})
and the observation of a real ``oscillatory'' pattern
(disappearance + reapperance). The first issue will be attacked by
the CERN-to-Gran Sasso experiment OPERA \cite{Schn}, and the
second  by new long-baseline or atmospheric experiments, probing
$\nu_\mu\to\nu_\mu$ disappearance with higher $L/E$ resolution
\cite{Tots,Schn,Taba}.

Let us now consider some current topics about the dominant
$\nu_\mu\to\nu_\tau$ channel and about subdominant effects due to
either extra states ($\nu_e$ or $\nu_s$) or new interactions.

\begin{figure}[t]
\vspace{+4truecm}
\hspace{-1.5truecm}
\epsfig{bbllx=0pt,bblly=0pt,bburx=500pt,bbury=400pt,width=8truecm,%
height=10truecm,figure=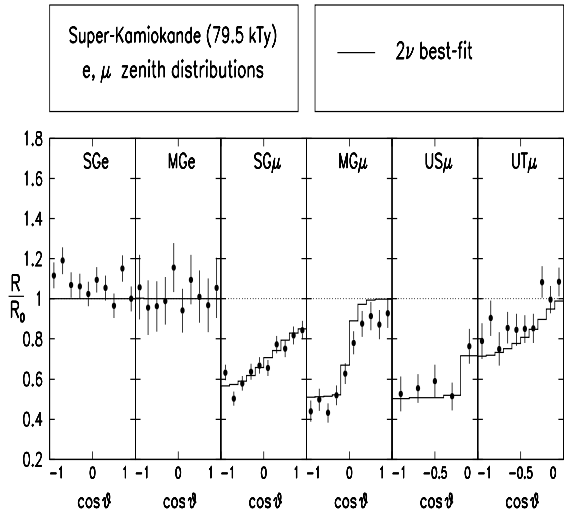} \vspace{-7.5truecm}
\caption{\footnotesize Best-fit predictions for normalized SK
lepton ($e,\mu$) rates $R/R_0$ for two-family $\nu_\mu\to\nu_\tau$
oscillations (solid histograms) in terms of the lepton zenith
angle ($\cos\theta$). The no-oscillation case corresponds to
$R/R_0=1$ (dotted line). Dots with error bars represent SK data
$\pm1\sigma_{stat}$ (SG=sub-GeV, MG=multi-GeV, US=Upward-stopping,
UT=Upward-throughgoing). See also \protect\cite{Subd} for
details.\vskip-6mm}
\end{figure}

\subsection{Dominant $\nu_\mu\to\nu_\tau$ oscillations}

The stability of $\nu_\mu\to\nu_\tau$ oscillations as the dominant
mechanism for atmospheric  $\nu_\mu$ disappearance is guaranteed
by several independent facts: (i) nonobservation of matter effects
associated to possible large $\nu_e$ or $\nu_s$ admixture
\cite{Tots}; (ii) strong upper bounds on additional $\nu_e$ mixing
from negative CHOOZ reactor searches \cite{CHOO,Scap}; (iii)
nonobservation of NC event suppression due to possible
$\nu_\mu\to\nu_s$ in SK \cite{CHOO}; (iv) hints of $\nu_\tau$
appearance in SK \cite{Tots}; (v) generally poorer fits provided
by alternative (``exotic'') explanations \cite{Tots}, with the
possible exception of a decoherence model \cite{Deco}.
 The stability of the best-fit
 parameters, on the other hand, is guaranteed by the
strong disappearance effect observed in the MG$\mu$ sample of SK
($E\sim {\rm few\ GeV}$, see Fig.~6): the onset of disappearance
at the horizon (fixing the length scale $L$) determines the
squared mass difference $m^2\sim E/L$, while the maximum
suppression  of the upward muon rate $(\sim 1/2)$ forces the
mixing amplitude $\sin^2 2\psi$ to be nearly maximal.

\begin{figure}[t]
\vspace{-1.9truecm}
\hspace{-1.2truecm}
\epsfig{bbllx=0pt,bblly=0pt,bburx=500pt,bbury=750pt,width=8.5truecm,%
figure=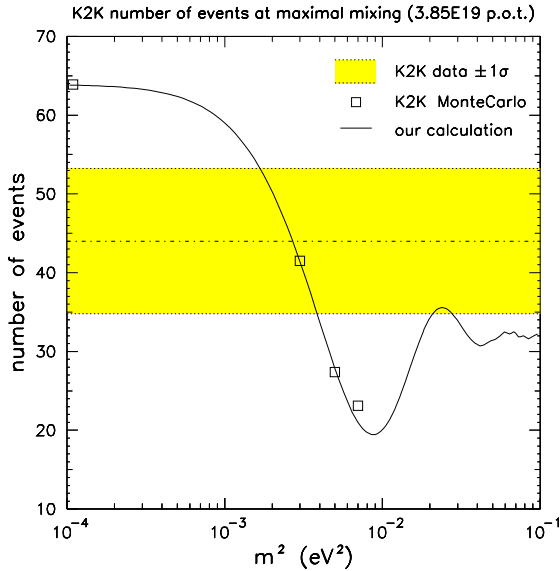} \vspace{-3.8truecm} \caption{\footnotesize K2K
number of events as a function of the squared mass difference
$m^2$ for maximal $\nu_\mu\to\nu_\tau$ mixing. Horizontal band:
K2K data \protect\cite{Hase}. Solid curve: Expectations in the
presence of oscillations \protect\cite{K2Ko}.\vskip-9mm. }
\end{figure}

\subsection{Impact of K2K}

\begin{figure}[t]
\vspace{-1.3truecm} \hspace{-1.8truecm}
\epsfig{bbllx=0pt,bblly=0pt,bburx=500pt,bbury=750pt,width=9truecm,%
figure=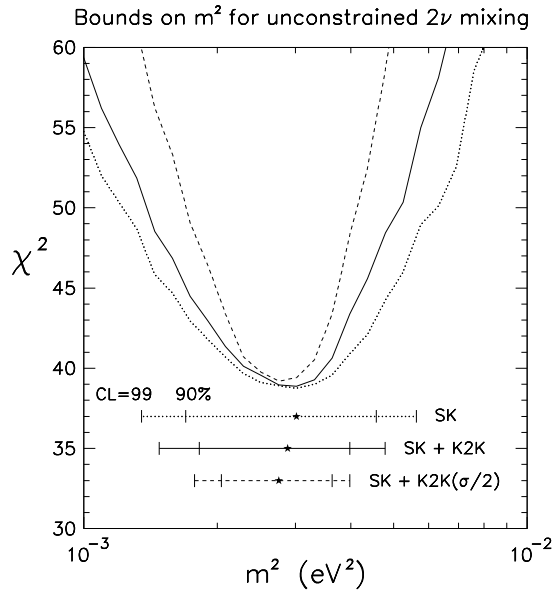} \vspace{-5.2truecm} \caption{\footnotesize Results
of the $\chi^2$ analysis of SK atmospheric and K2K accelerator
data in terms of the leading mass difference $m^2$ (for
unconstrained $2\nu$ mixing). Increasingly tighter bounds are
obtained by adding to SK data the K2K data, and eventually by
halving the K2K errors \protect\cite{K2Ko}.\vskip-4mm}
\end{figure}

The KEK-to-Kamioka (K2K) long-baseline $\nu$ experiment ($L=250$
km) aims to verify the SK atmospheric $\nu$ anomaly through
disappearance of low-energy ($>1$ GeV) accelerator $\nu_\mu$'s.
The recent K2K data \cite{Hase} (44 events observed against 64
expected) already suggest that the null hypothesis of no
oscillations is rejected at 97\% C.L. \cite{Hase}.

One can make a further step and analyze the consistency and the
impact of  K2K data in the context of the $\nu_\mu\to\nu_\tau$
oscillation scenario \cite{K2Ko}.

Figure~7 shows the number of events expected in K2K (solid curve)
as a function of $m^2$ (for maximal mixing, $\tan^2\psi=1$), as
compared with the K2K data  ($1\sigma$ allowed band). The
theoretical curve crosses the central value of the band at
$m^2\sim 2.5\times 10^{-3}$ eV$^2$, in  agreement with the value
derived from SK atmospheric $\nu$ data. Therefore, K2K supports
the same $\nu_\mu\to\nu_\tau$ oscillation scenario favored by
atmospheric neutrino experiments, and it makes sense to combine SK
and K2K data in a global analysis (dominated, of course, by the
more accurate SK data).

The results are shown in Fig.~8, in terms of bounds on $m^2$ from
SK only, from SK+K2K, and from SK+K2K with halved K2K errors. It
can be seen that K2K can appreciably tighten both the lower and
the upper bound on $m^2$. Strengthening the lower bound is
particularly important for the new generation of accelerator
experiments with longer baselines \cite{Schn}, which would rapidly
loose sensitivity to oscillation effects for $m^2$ approaching
$\sim 1\times 10^{-3}$ eV$^2$.

\begin{figure*}[t]
\vspace{-2.8truecm}
\hspace{-1.6truecm}
\epsfig{bbllx=0pt,bblly=0pt,bburx=500pt,bbury=750pt,width=15.5truecm,%
height=16truecm,figure=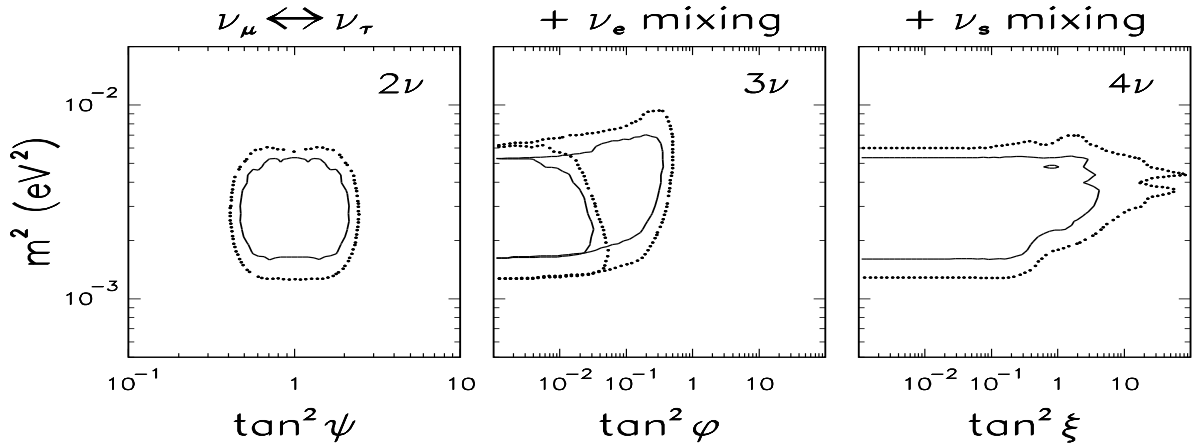} \vspace{-8.truecm}
\caption{\footnotesize Bounds on atmospheric neutrino oscillation
parameters $(m^2,\tan^2\psi)$ for dominant flavor states
$\nu_\mu,~\nu_\tau$ (left panel), and upper bounds on additional
mixing with extra states, parametrized through $\tan^2\phi$ for
$\nu_e$ (middle panel, with and without CHOOZ), and through
$\tan^2\xi$ for $\nu_s$ (right panel). The middle and left panels
are relevant for scenarios with three and four neutrino mixing
\protect\cite{Subd}.}
\end{figure*}

\subsection{Subdominant oscillations from extra states}

It is natural to assume that at least one extra state ($\nu_e$)
can play a role besides dominant $\nu_\mu\to\nu_\tau$ oscillations
 ($3\nu$ mixing). A possible light sterile state $\nu_s$ might
also participate to oscillations in extended $4\nu$ scenarios
\cite{Giun}. In both cases, the standard $L/E$ dependence of the
$\nu_\mu\to\nu_\tau$ oscillation phase is modified by the
so-called Mikheyev-Smirnov-Wolfenstein (MSW) matter effects in the
Earth. Simbolically,
$$
{\rm Oscillation \ phase} \propto [L/E] \oplus [{\rm MSW}]\ .
$$
Nonobservation of deviations from the standard $L/E$ dependence in
SK places significant constraints on additional mixing of
$\nu_{\mu,\tau}$ with additional flavor states. The results are
shown in Fig.~9, as we now discuss.

\subsubsection{Additional $\nu_e$ mixing}

The middle panel of Fig.~9 shows the bounds on additional $\nu_e$
mixing, expressed in terms of the leading squared mass difference
$m^2$ and of $\tan^2\phi$, where $\phi=\theta_{13}$ in usual
$3\nu$ mixing notation. The looser bounds  in the panel arise from
the absence of observed MSW effects and of distortions of the
electron samples in SK (see also Fig.~6). The tighter bounds refer
to the SK+CHOOZ combination, dominated by the CHOOZ nonobservation
of $\bar\nu_e$ disappearance.

Altogether, SK+CHOOZ set an upper bound of a few \% on additional
$\nu_e$ mixing. Such bound is probably too tight to allow
detection of residual matter effects with higher SK statistics
\cite{Subd,Pere}. Establishing a nonzero value for
$\phi=\theta_{13}$ is a major challenge for future neutrino
oscillation experiments, with far-reaching consequences also for
non-oscillatory probes of the neutrino mass and mixing \cite{Klap}
and for model building and leptonic CP violation \cite{Lang}.

\subsubsection{Additional $\nu_s$ mixing}

In this case, matter effects are typically somewhat weaker than
for $\nu_e$ mixing, and there is no analogous of the CHOOZ
experiment. Therefore, the upper bounds on $\nu_s$ mixing
(parametrized in Fig.~9 in terms of a suitable angle $\xi$, see
\cite{Subd}), are weaker than the previous ones for $\nu_e$
mixing.

However, it should be added that the limits in the right panel of
Fig.~9 can be improved by adding: further SK data on NC enriched
events \cite{Tots}; SK hints for $\tau$ appearance \cite{Tots};
MACRO muon data \cite{Scap,Vall}. Our educated guess is that a
global combination of all such data would limit subdominant
$\nu_s$ mixing below 20--30\%.

\begin{figure}[t]
\vspace{-.3truecm}
\hspace{-1.4truecm}
\epsfig{bbllx=0pt,bblly=0pt,bburx=500pt,bbury=750pt,width=8truecm,%
figure=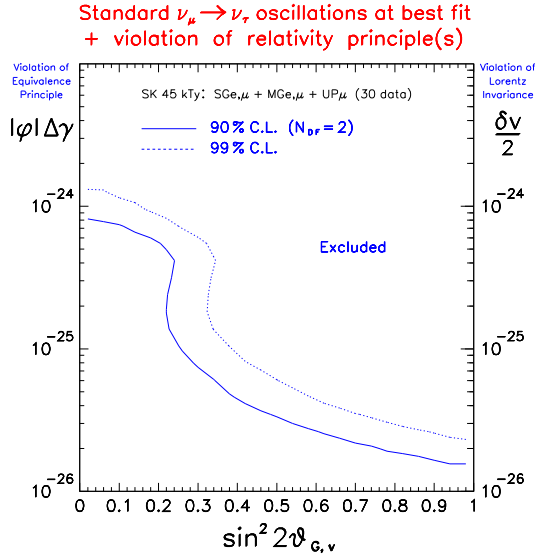} \vspace{-4.7truecm} \caption{\footnotesize
Standard $\nu_\mu\to\nu_\tau$ mass-mixing oscillations plus
violations of relativity principles: Upper bounds from SK data.
See \protect\cite{Viol} for details.\vskip-6mm}
\end{figure}

\subsection{Subdominant oscillations from extra interactions}

The standard $L/E$ oscillation phase of atmospheric
$\nu_\mu\to\nu_\tau$ oscillations  can be modified not only by
extra states, but also by extra (nonstandard) interaction of
$\nu_{\mu,\tau}$ with a ``generalized'' background, which can be
either the Earth matter, or the local gravitational field, or the
space-time itself. A large class of nonstandard interactions
predict extra oscillations phases with a power-law dependence on
energy \cite{Viol}
$$
{\rm Oscillation \ phase} \propto [L/E] \oplus [L/E^n]\ .
$$
with integer $n$. The broad $L/E$ range spanned by SK sets severe
limits on the amplitude of nonstandard $(n\neq -1)$ phases, as we
show for two representative cases. Future (very) long baseline
accelerator experiments will not easily improve such limits
\cite{Zuka}, due to the their much narrower range of $L/E$ probed.

\begin{figure}[t]
\vspace{-.8truecm} \hspace{-.5truecm}
\epsfig{bbllx=0pt,bblly=0pt,bburx=500pt,bbury=750pt,width=6.5truecm,%
figure=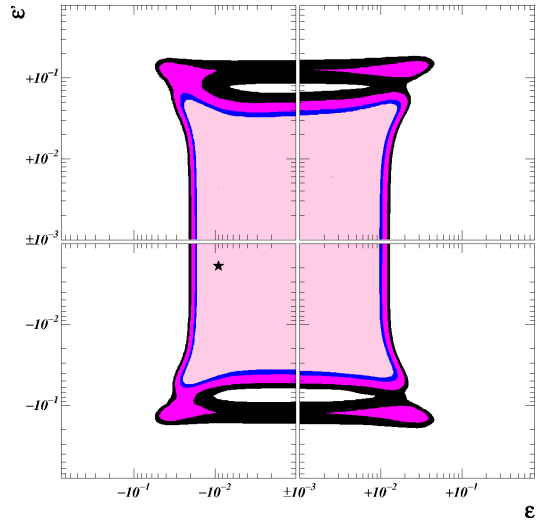} \vspace{-2.1truecm} \caption{\footnotesize
Standard $\nu_\mu\to\nu_\tau$ mass-mixing oscillations plus extra
four-fermion interactions: Upper bounds from SK data. See
\protect\cite{Forn} for details. \vskip-6mm}
\end{figure}

\subsubsection{Additional VEP, VLI}

Violations of the equivalence principle (VEP), i.e., different
coupling of $\nu_{\mu\tau}$ to gravity, can generate flavor
oscillations.  Violations of Lorentz invariance (VLI), i.e.,
different limiting speeds for $\nu_{\mu,\tau}$ can also generate
oscillations. Both cases generate extra oscillation phases with
$n=+1$ (phase $\propto L\cdot E$) \cite{Viol}. When added on top
of standard $\nu_\mu\to\nu_\tau$ transitions, such hypothetical
violations are found to worsen the fit to SK data. The lack of
evidence in favor of them allows to set strong upper bounds to
their amplitude, as shown in Fig.~10 \cite{Viol}. Such bounds are
much tighter than the corresponding ones derived in another
``oscillation laboratory'', namely, the kaon system.

\subsubsection{Additional FCNC, FDNC}

Additional NC-type, four-fermion interactions can arise in SUSY
models with R-parity breaking, leading to flavor-changing (FC) or
flavor-diagonal (FD) neutrino transitions in matter. The
amplitudes of such interactions are usually parameterized as
``fractions'' ($\epsilon$'s) of the standard Fermi amplitude
$G_F$. Their effect on oscillations is to produce
energy-independent $(n=0)$ extra phases. No evidence is found for
such phases, and upper bounds on the $\epsilon$'s have been
recently derived in \cite{Forn}. Figure~11 reports the basic
results of \cite{Forn}: nonstandard interactions cannot exceed a
few percent strength (in units of $G_F$).

\section{COMMENTS ON $4\nu$ MIXING}

While solar, atmospheric, and CHOOZ data can be accommodated in a
scenario involving only the three known neutrinos
\cite{Lang,Thre}, problems arise when one tries to embed also the
controversial evidence for oscillations in the $\nu_\mu\to\nu_e$
channel found by LSND \cite{LSND}, which is not supported (but not
totally excluded) by the competing Karmen experiment (see
Fig.~12). The evidence is characterized by small mixing and a
large squared mass difference, which is incompatible with solar
and atmospheric neutrino parameters, unless a fourth (sterile)
neutrino state is introduced \cite{Giun}.

Two four-neutrino mass spectra can be envisaged : one with three
close states and a lone state (3+1) \cite{Smir}, and one with two
separated doublets (2+2) \cite{Giun}. Such spectra can, in
principle, accommodate the three independent square mass
differences needed to drive solar, atmospheric, and LSND
oscillations. However, mixing angles turn out to be difficult to
arrange without spoiling the agreement with some data.

In the 2+2 case, it turns out that the fractional admixture of
$\nu_s$ in the ``solar''  and ``atmospheric'' doublets must add up
to one. Symbolically:
$$
2+2:\ [\nu_s]_{\rm sol}+[\nu_s]_{\rm atm} = 1\ .
$$
However, the best fits to $[\nu_s]_{\rm sol}$ and $[\nu_s]_{\rm
atm}$ are close to zero (as we have seen in Sec.~2.3 and 3.3.2,
respectively), so that the above sum rule cannot be satisfied
unless  either $[\nu_s]_{\rm sol}$ or $[\nu_s]_{\rm atm}$ are
pushed to their upper limits at 2--3$\sigma$ \cite{Pena}.

In the 3+1 case, it turns out that the LSND appearance probability
$P_{\mu e}$ is proportional to the $\nu_\mu$ disappearance
probability (constrained by the CDHSW accelerator experiment) and
to the $\nu_e$ disappearance probability (constrained by the Bugey
reactor experiment). Symbolically:
$$
3+1: \ P_{\mu e}^{\rm LSND} \propto P_{\mu\mu}^{\rm CDHSW}\cdot
P_{ee}^{\rm Bugey}\ .
$$
Since both factors on the r.h.s.\ are experimentally consistent
with zero, the expected LSND probability is doubly suppressed, and
it turns out to be too small to fit  the observed data
\cite{Vall}.

In conclusion, the LSND evidence does not seem to fit well in
$4\nu$ schemes: tension is generated between solar and atmospheric
data in the 2+2 case, and between accelerator and reactor data in
the 3+1 case. This tension is not strong enough, however, to
exclude $4\nu$ mixing with high confidence yet  \cite{Vall}.

In any case, independently from the above (model-dependent)
arguments, the LSND issue will be soon settled by the MiniBoone
experiment \cite{Stef}. Disconfirmation of LSND would reinforce
the standard $3\nu$ oscillation interpretation of solar and
atmospheric $\nu$ data. A confirmation would instead imply a
serious global re-examination of the oscillation evidence, in
order to find new ways to make LSND compatible with world data.

\begin{figure}[t]
\vspace{-.9truecm} \hspace{-.2truecm}
\psfig{width=8.5truecm,%
file=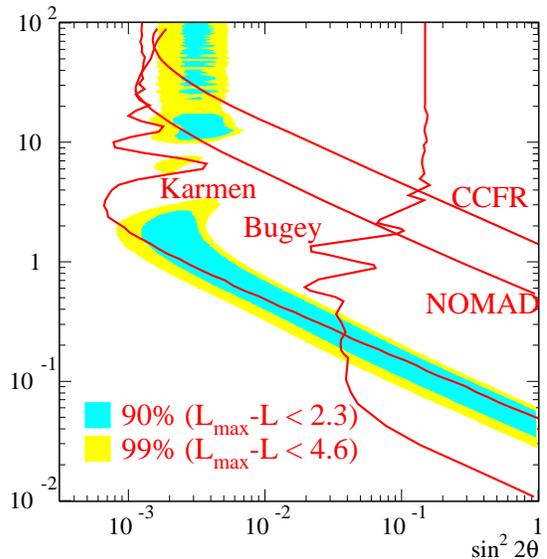} \vspace{-1.5truecm} \caption{\footnotesize LSND
allowed region (shaded) {\em vs\/} regions excluded by other
experiments in the $\nu_\mu\to\nu_e$ mass-mixing plane.\vskip-7mm}
\end{figure}

\section{SUMMARY}

We have reviewed recent topics in the neutrino oscillation
phenomenology. The solar neutrino evidence for active $\nu$
oscillations is significantly strengthened by SNO data as compared
to SK data, independently on specific solar or oscillation models.
The post-SNO oscillation fit strongly favors large-mixing
solutions, with a preference for relatively high values of the
neutrino squared mass difference (LMA solution). Solar neutrinos
do not provide evidence for a sterile $\nu$ admixture, although a
subdominant $\nu_s$ component can be easily tolerated. Atmospheric
neutrinos are beautifully explained by dominant
$\nu_\mu\to\nu_\tau$ oscillations. Preliminary K2K data are
consistent with this interpretation. Upper bounds can be placed on
subdominant admixtures of extra states ($\nu_e$ or $\nu_\tau$) and
on subdominant contributions of extra interactions (beyond the
standard model). Solar and atmospheric neutrino data can be easily
embedded in a $3\nu$ mixing scenario. However, when one tries to
add a fourth sterile neutrino to embed also the LSND data, tension
arises between subsets of data. A series of new experiments will
greatly improve and refine the current picture of neutrino mass
and mixing in the near future (or in the next decade at most), and
will possibly set new challenges for the phenomenological
interpretation and for model building.

\section*{Acknowledgments}
This work is supported by the Italian INFN and MIUR under the
``Astroparticle Physics'' project.

\end{document}